\documentclass[a4paper, twocolumn, unpublished, noarxiv, 10pt]{quantumoriginal}
\pdfoutput=1

\usepackage[utf8]{inputenc}
\usepackage[english]{babel}
\usepackage[T1]{fontenc}
\usepackage{amsmath}
\usepackage{hyperref}
\usepackage{braket}
\usepackage{caption}
\usepackage{subcaption}
\usepackage{adjustbox}
\usepackage{multirow}
\usepackage{makecell}

\DeclareMathOperator{\Tr}{Tr}

\begin{document}

\title{Infinite Distance Extrapolation: How error mitigation can enhance quantum error correction}

\author{George Umbrarescu}
\affiliation{Department of Physics and Astronomy, University College London, WC1E 6BT London, United Kingdom}
\orcid{0000-0003-4376-2147}
\email{george.umbrarescu.20@ucl.ac.uk}
\author{Oscar Higgott}
\affiliation{Department of Physics and Astronomy, University College London, WC1E 6BT London, United Kingdom}
\orcid{0000-0001-9880-5218}
\author{Dan E. Browne}
\affiliation{Department of Physics and Astronomy, University College London, WC1E 6BT London, United Kingdom}
\orcid{0000-0003-3001-158X}

\maketitle

\begin{abstract}
    Quantum error mitigation (QEM) and quantum error correction (QEC) are two research areas that are often considered as distinct entities, and the problem of combining the two approaches in a non-trivial way has only recently started to be explored. In this paper, we explore a paradigm at the intersection of the two, based on the error mitigation technique of Zero-Noise Extrapolation (ZNE), that uses the distance of an error correcting code as a noise parameter. This is distinct from some alternative approaches, as QEC is here used as a subroutine inside the QEM framework, while other proposals use QEM as a subroutine inside QEC experiments. Intuitively, we exploit the fact that a reduction in the physical noise level is analogous to an increase in the code distance, as both of them result in a decrease in the logical error rate. As such, the extrapolation to zero noise in the case of ZNE becomes comparable to the extrapolation to infinite distance in the case of this method. We describe how to calculate expectation values from a fault-tolerant computation, and we gain some analytical intuition for our ansatz choice. We explore the performance of the considered method to reduce the errors in a range of expectation values for a realistic circuit-level noise model and realistic device imperfections on the rotated surface code, and we particularly show that the performance of the method holds even in the case of non-stabiliser input states.
\end{abstract}

\section{Introduction}

Quantum computers have the potential to solve difficult problems in fields such as chemistry \cite{mcardle2020quantum, bauer2020quantum}, optimisation \cite{abbas2024challenges} and finance \cite{herman2022survey}. However, the implementation of such algorithms on real quantum devices is plagued with noise and control imperfections that degrade the quality of the results obtained. Two of the leading approaches for reducing these errors are QEC and QEM.

QEC was the first approach to be introduced and was inspired by error correction in classical computers and communications. It relies on encoding the system of interest into a larger system such that local perturbations do not have global effects on the information stored in a delocalised fashion \cite{aharonov1999fault, terhal2015quantum}. While QEC can suppress errors to arbitrarily small levels, it requires physical resources that are believed to become available only in the FTQC era.

QEM, on the other hand, was proposed as a solution for reducing errors in the NISQ era, where we do not have sizeable devices with low error rates, but we have the ability to trade spatial resources, in the form of more physical qubits, for temporal ones, in the form of more executions and measurements of the circuit. By post-processing the results obtained from the execution of different variants of the original circuits, we can reduce the bias in our estimate of an expectation value at the cost of increasing the variance \cite{cai2021practical, cai2023quantum}.

Of particular relevance nowadays might be the Early Fault-Tolerant (EFT) era \cite{katabarwa2024early}, where devices are getting bigger in size and the error rates are getting better, but full fault-tolerant algorithms are still not possible to be implemented. As such, there is great scope for combining QEC and QEM approaches in a non-trivial way to bridge this gap between heuristic post-processing and rigorous error correction. It is especially interesting to investigate in what circumstances can QEM enhance QEC when under the pseudothreshold of the code. 

The literature on bringing QEC and QEM together is gaining interest, with several proposals over the past years: in \cite{piveteau2021error}, the authors present a method to improve the quality of injecting magic $T$ states through quasi-probability decomposition, in \cite{suzuki2022quantum}, Pauli recovery operators for probabilistic error cancellation are applied virtually instead of physically, in \cite{tsubouchi2023virtual}, symmetry expansion is generalised to virtual quantum error detection, which can reduce the need to physically measure all the stabiliser generators, in \cite{gonzales2023fault}, Pauli Check sandwiching is shown to be related to quantum error detection, and in \cite{liu2025virtual}, virtual state purification is generalised to work in an error correction setting. Even more recently, in \cite{xie2026noise}, a circuit-level variant of PEC inversion is used as a subroutine inside QEC, in \cite{zhou2025error, dincua2025error, aharonov2025syndrome}, the confidence of the decoder is used to inform further post-processing through error mitigation, and in \cite{jeon2026quantum}, it is proven that by integrating a linear QEM method such as PEC at the physical layer, some of the leading order error terms are suppressed to the benefit of a method-agnostic decoder.

Zero-noise extrapolation is one of the original QEM schemes that have been proposed \cite{temme2017error, endo2018practical}. The principle behind it is that noisy expectation values of a certain observable can be measured for the original noise level of a quantum computer, as well as for amplified noise levels on the same device. Then, a function is chosen to be fitted through the data points given by the pairs of noise levels and their respective noisy expectation values. The function is evaluated for a parameter equal to 0 for the noise level, which corresponds to the noiseless expectation value. This method is simple, yet powerful, and it should give an improvement compared to the original estimate of the noisy expectation value as long as the scaling of the noise is well-behaved enough for our chosen functional form.

The intuition behind adapting zero-noise extrapolation to an error correction setting is as follows. In QEC, one parameter that we can control in our experiments is the code distance, which can be changed purely by increasing the number of used physical qubits. By the threshold theorem, when the physical error rate is lower than the threshold error rate, increasing the distance will decrease the logical error rate (LER). Therefore, infinitely increasing the distance will lead to obtaining a LER equal to zero, which means that all the errors are detected, decoded and corrected. This can be viewed as similar to having no noise at the logical level. As such, increasing the noise artificially and then extrapolating to zero noise in the pure QEM picture is analogous to decreasing the code distance and then extrapolating to the infinite distance limit in the QEC picture, as decreasing the noise at the physical level is analogous to increasing the code distance. The QEC setting is particularly appropriate for this extrapolation technique because the dependence of expectation values on the code distance has been shown to be well-behaved \cite{fowler2012surface}, making it a more well-controlled application than a general NISQ circuit that uses ZNE. 

Practically, every fault-tolerant component of the computation is rescaled by the distance of the code used, and we run experiments with several different distances. Implementing logical gates requires protecting the logical operators from both time-like and space-like errors, as the computation progresses in time and space. For both memory and stability components \cite{gidney2022stability}, the distance affects the computation both spatially, in terms of the number of physical qubits encoding the logical qubits, and temporally, as the number of stabiliser measurement rounds also scales with the distance. We expect the scaling of the decrease in the logical error rate of a stability component as a function of the rounds of stabiliser measurements to be similar to the one for a memory component as a function of the code's spatial distance. As such, we expect the LER for a lattice surgery operation required to implement a logical gate to scale in a similar way with the distance, since it has itself several memory and stability components. Therefore, in the target experiments, all the components of the logical computation are scaled by the distance of the code both spatially and temporally, allowing us to obtain a relationship between the EV of the full computation and the distance.

In particular, the idea of using the distance as a scaling parameter for expectation value extrapolation was first introduced in \cite{wahl2023zero} and experimentally demonstrated for the repetition code in \cite{zhang2025demonstrating}. In \cite{wahl2023zero}, this approach is tested for the case of logical randomised benchmarking circuits, in conjunction with canonical ZNE folding-based error amplification. However, that paper did not implement a full quantum memory experiment including decoding, but rather used a simple model which mimicked the effect of error correction via an effective Pauli channel of a QEC cycle. It therefore remained an open question whether the extrapolation approach would be effective in a more accurate simulation, or indeed on a device itself.

The paper is structured as follows. Section \ref{sec:background} introduces the surface code, the noise models, and the relevant QEM background. Section \ref{sec:ide} formulates IDE and motivates the logical-error-rate ansatz. Section \ref{sec:bypassing_methods} describes how we obtain non-trivial expectation values efficiently despite the presence of non-Clifford structure. Section \ref{sec:results} presents the numerical results, including performance across different input states, robustness to spatial noise variation, and resource estimates. We conclude in Section \ref{sec:conclusion} with the regime of applicability of IDE and directions for future work.

\section{Background}
\label{sec:background}

\subsection{Noise in quantum processors}

\paragraph{General noise}

Noise in quantum systems arises from the interaction with the environment (incoherent, stochastic noise) or from imperfect experimental control (coherent noise) \cite{wallman2015estimating}. Stochastic noise is non-unitary and leads to an increase in entropy through information leakage into the environment, but only incurs errors that scale linearly in the depth of the circuit. Coherent noise is unitary, therefore preserving the purity of the state, but it accumulates quadratically in the depth of the circuit. \cite{wallman2016noise}

\paragraph{Pauli noise}

Due to the Pauli Twirling Approximation \cite{geller2013efficient}, we can assume that all the noise channels in the system (both stochastic and coherent) can be transformed into stochastic Pauli channels. This allows us to work with only a discrete set of errors, which is advantageous for QEC.

\subsection{Quantum error correction (QEC)}

\subsubsection{Theory}

\paragraph{Quantum error correcting codes}

QEC codes are defined by their parameters $[[n, k, d]]$, where $n$ is the number of physical qubits, $k$ is the number of logical (encoded) qubits, and $d$ is the distance of the code, such that the number of detectable errors is $d - 1$ and the number of correctable errors is $\lfloor (d - 1) / 2 \rfloor$ \cite{terhal2015quantum}.

\paragraph{Detector error model (DEM)}

A detector error model \cite{gidney2021fault, higgott2025sparse, derks2024designing} is a description of the physical error mechanisms that can flip the detectors and the logical observables of the circuit. The detectors of the circuit are linear combinations of measurement outcomes that are deterministic in the noiseless case. The DEM corresponds to the Tanner/decoding graph.

\paragraph{Decoding}

Through decoding, we aim to identify that an error has occurred, and how the error can be corrected. By running a decoder on the DEM, we can determine if our observable has been flipped (outputted a measurement result inconsistent with the noiseless simulation) in the presence of noise. Minimum weight perfect matching (MWPM) is a widely popular decoding algorithm for the surface code \cite{dennis2002topological, fowler2012surface, higgott2022pymatching, higgott2025sparse}. It attempts to find the graphlike physical error with the lowest weight that could have produced the syndrome. 

Once we select a code and a decoder, we can calculate the threshold of the code family, which is the physical error rate below which increasing the distance of the code reduces the logical error rate.

\paragraph{Fault-tolerance and code distance}

Fault-tolerance is a property of a circuit such that an individual faulty component does not propagate into an overall logical failure \cite{aharonov1999fault}. The notion of distance for a code can be generalised to the level of a circuit, where the parity checks cannot be implemented as direct coherent measurements, and their circuit implementation is also faulty. A fault-tolerant circuit aims to preserve the maximal circuit distance $d$. Equivalently, the distance is given by the minimum weight of a logical operator (minimum number of errors that could lead to a logical error) \cite{terhal2015quantum}.

\subsubsection{Surface code}

\begin{figure}
    \centering
    \includegraphics[width=1.0\linewidth]{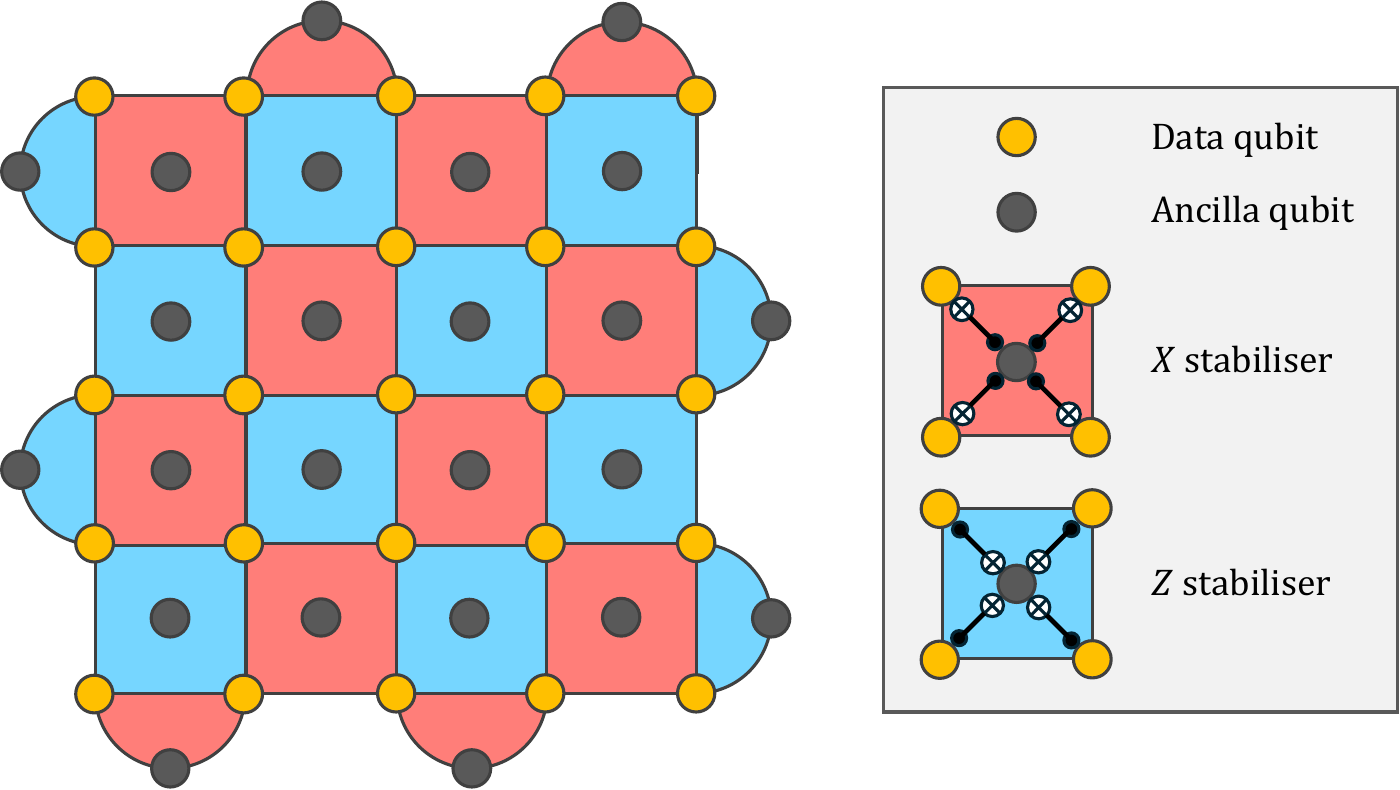}
    \caption{Rotated surface code for $d = 5$.}
    \label{fig:surface_code}
\end{figure}

The surface code \cite{dennis2002topological, fowler2012surface} is one of the most promising candidates for the practical implementation of QEC in the FT era. In this paper, we consider the rotated version of the surface code \cite{horsman2012surface}, which provides significant savings in the number of data qubits compared to the unrotated version ($d^2$ vs $2d^2$). A pictorial representation is shown in Figure \ref{fig:surface_code}.

\paragraph{Stabilisers}

The $Z$ and $X$ plaquette operators are defined in the usual way, with the stabilisers having support on 4 data qubits in the bulk and 2 data qubits at the boundaries.

\paragraph{Logical operators}

In Figure \ref{fig:surface_code}, the logical $\overline{Z}$ operators are defined horizontally, while the logical $\overline{X}$ operators are defined vertically.

\paragraph{State preparation and measurement}

Magic state preparation and magic measurements rely on direct stabiliser measurements via Pauli product measurements, without considering any limitations in connectivity or measurement capacity of the hardware. These can easily be done in software with a stabiliser simulator, eliminating the need to devise fault-tolerant schemes for initialisation and measurement.

When considering a realistic setting, in the surface code, $Z$- and $X$-basis initialisations and measurements can be done transversally, but this is not the case for the $Y$ basis. Recently, it has been shown how to implement inplace $Y$-basis state initialisation and measurement \cite{gidney2024inplace}.

Therefore, using these tools, we can implement the logical $\ket{0}$ and $\ket{1}$ states in the $Z$ basis, the logical $\ket{+}$ and $\ket{-}$ states in the $X$ basis and the logical $\ket{+i}$ and $\ket{-i}$ states in the $Y$ basis.

For our experiments, we consider a realistic memory experiment that uses $O(d)$ rounds of syndrome extraction, in an attempt to mitigate the effects of measurement errors on the misreporting of stabiliser measurements \cite{cain2025fast}.

\subsubsection{Estimating logical error rates}

Logical error rates are defined based on the proportion of flipped observable results, when decoding fails to find an appropriate correction:
\begin{equation}
    P_L = \frac{n_\text{fails}}{n_\text{shots}}
\end{equation}

\subsubsection{Estimating expectation values}

For a trivial expectation value of $\pm 1$, such as the ones obtained in memory experiments which amount to logical identity, the expectation value is straightforwardly related to the logical error rate:
\begin{equation}
    E = \pm\left( 1 - 2 \times P_L \right)
\end{equation}
In this case, for a single logical qubit, the state is prepared and measured in the same basis, e.g. measuring $\braket{\overline{Z}}$ on $\ket{0}_L$ or $\braket{\overline{X}}$ on $\ket{+}_L$. The DEM used for decoding is obtained directly from the circuit used for state preparation.
More general and non-trivial expectation values that are in between $-1$ and $+1$ (from non-Clifford logical operations) cannot be directly estimated efficiently using a stabiliser simulator. 

In addition, we want to highlight the important point that the DEM used for decoding needs to be obtained from the circuit implementing state preparation in the same bases as the required measurement, otherwise there is the risk of producing non-deterministic detector outcomes. For example, preparing the $\ket{0}_L$ state and wanting to measure $\braket{\overline{X}}$ requires that the DEM used for decoding is produced from the circuit preparing the $\ket{+}_L$ state and measuring $\braket{\overline{X}}$.

We believe this subtle point is relevant as we go beyond testing the performance of QEC codes by simply measuring LERs, and we move onto estimating real expectation values in fault-tolerant computations.

\subsection{Quantum error mitigation (QEM)}

In the field of QEM, the goal of the proposed methods is to reduce the bias in the estimator of the expectation value, at the expense of increasing its variance \cite{cai2021practical}. This allows us to obtain the mitigated expectation value, $E^*$, which is closer to the noiseless expectation value.

\subsubsection{Zero-noise extrapolation (ZNE)}

For a noise level quantified by the parameter $\lambda$, we can calculate the noisy expectation value of an observable $O$ on a noisy state $\rho(\lambda)$ as follows:
\begin{equation}
    E_\lambda = \braket{O}_\lambda = \Tr{[\rho(\lambda) O]}
\end{equation}
The idea behind ZNE is to estimate $E_\lambda$ for a sequence of values of $\lambda$ to obtain a sequence of pairs $\{(\lambda_1, E_{\lambda_1}), (\lambda_2, E_{\lambda_2}), ..., (\lambda_k, E_{\lambda_k})\}$. Then, we fit the parameters $\{c_1, c_2, ..., c_m\}$ of a function $f(\lambda, c_1, c_2, ..., c_m)$ such that:
\begin{equation} \label{eq:E_lambda}
    E_\lambda \approx f(\lambda, c_1, c_2, ..., c_m)
\end{equation}
The error-mitigated expectation value is obtained by evaluating the fitted function at the point $\lambda = 0$, equivalent to the limit of no noise:
\begin{equation}
    E^* = \lim_{\lambda \to 0} f(\lambda, c_1, c_2, ..., c_m)
\end{equation}

For ZNE, various functional forms have been proposed, such as linear, quadratic, higher-order polynomials, exponentials and combinations thereof \cite{giurgica2020digital}.

Some of the limitations of ZNE are that we need a good way to quantify the level of noise in the circuit, and more crucially a reliable way to amplify the noise without losing too much signal.

\subsubsection{Richardson extrapolation}

One particularly elegant and commonly used method for the fitting procedure is called Richardson extrapolation \cite{temme2017error}.
Here, the polynomial degree is chosen to be one less than the total number of data points ($m$), so the ansatz has $m$ coefficients \cite{giurgica2020digital}:
\begin{equation} \label{eq:richardson_ansatz}
    E_\text{Rich}(\lambda) = c_0 + c_1 \lambda + ... + c_{m - 1} \lambda^{m - 1}
\end{equation}
To obtain the extrapolated value in the limit of zero noise, we have to use interpolating Lagrange polynomials, giving:
\begin{equation} \label{eq:richardson_yk}
    E_\text{Rich}(0) = c_0 = \sum_{k = 1}^m y_k \prod_{i \neq k} \frac{\lambda_i}{\lambda_i - \lambda_k}
\end{equation}
where $\{y_k\}$ are the data points, in our case the expectation values $\{E_{\lambda_k}\}$. In the traditional ZNE setting, $\{\lambda_k\}$ are the quantified noise levels, for example given by the number of repetitions of the circuit in the folding procedure used to amplify the original noise level.

\section{Infinite distance extrapolation (IDE)}
\label{sec:ide}

As described in the introduction, we adapt ZNE to the QEC setting. Specifically, by using the equivalence between reducing the noise at the physical level and increasing the distance at the logical level, we now perform an extrapolation to the infinite distance limit, rather than the zero noise limit. Varying the distance is a more controlled method of acquiring data for fitting the ansatz, and each additional point at higher distance has a better signal-to-noise ratio, while for traditional ZNE, artificially increasing the noise makes the new EV estimates more unreliable.

In the case of IDE, we estimate $E_d$ for a sequence of distances $\{d_k\}$ to obtain a sequence of pairs $\{(d_1, E_{d_1}), (d_2, E_{d_2}), ..., (d_k, E_{d_k})\}$. Then, we fit the parameters $\{c'_1, c'_2, ..., c'_m\}$ of a function where $d$ is analogous to the noise parameter $\lambda$ in Eq. \eqref{eq:E_lambda}, such that:
\begin{equation}
    E_d \approx f'(d, c'_1, c'_2, ..., c'_m)
\end{equation}
and:
\begin{equation} \label{eq:f_d}
    E^* = \lim_{d \to \infty} f'(d, c'_1, c'_2, ..., c'_m)
\end{equation}
where $E^*$ is the mitigated expectation value.

In the remainder of this section, we provide the justification of the LER ansatz used in IDE, we adapt the Richardson procedure to this new setting, and we give further details about the implementation of the IDE method on hardware.

\subsection{Analytics of the LER ansatz}

We are now going to start from previously-derived formulas for the logical error rate in the surface code and obtain the ansatz used for fitting expectation values in IDE.

Analytical expressions for the logical error rate for various noise models can be obtained by considering the possible error mechanisms that give rise to logical errors. By counting the possible error paths, we can estimate and weigh the contributions of different classes of noisy operations, such as 1-qubit gates, 2-qubit gates, measurements and idling. One of the seminal works to do this is Ref. \cite{fowler2012surface}, which uses these statistical arguments to obtain that the logical error rate depends on the (odd) distance and the physical error rate as such:
\begin{equation}
    P_L \sim d \frac{d!}{(d_e - 1)!d_e!} p^{d_e}
\end{equation}
where $d_e = (d + 1) / 2$. Based on this, the logical error rate is dominated by one term as:
\begin{equation}
    P_L \sim p^{d_e}
\end{equation}
Empirically, for their considered noise model, Ref. \cite{fowler2012surface} then shows that the LER could take the form:
\begin{equation}
    P_L \approx 0.03 (p / p_{th})^{d_e}
\end{equation}
below the threshold $p_{th}$. Using the identity $p^d = e^{(\ln{p}) \cdot d}$ and based on the previous arguments, for the logical error rate in our work, we assume an expression of the form:
\begin{equation}
    P_L(d) = B e^{-C d}
\end{equation}

We can then simply write the noisy expectation values as a sum of two terms: a noiseless bias part, $A$, which is the mitigated expectation value $E^*$ (Eq. \eqref{eq:f_d}), and an error part, $B e^{-Cd}$, which incorporates the effect of noise and vanishes in the limit of infinite distance (zero noise). This proposed ansatz thus has the form:
\begin{equation}
\label{eq:ansatz_1d}
    y(d) = A + B e^{-C d}
\end{equation}
We can additionally try to incorporate more complex effects in the proposed ansatz. For example, assuming errors in different Pauli bases contribute separately to the overall logical error scaling, we can assimilate their effect as an additional exponential in the ansatz. For example, a possible ansatz has the form:
\begin{equation}
    y(d) = A + B_{X/Z} e^{-C_{X/Z} d} + B_Y e^{-C_Y d}
\end{equation}
To support this, in \cite{gidney2024inplace} it was shown that the $Y$ basis LER in the surface code is higher than for the $X$ and $Z$ bases, even in the case of fault-tolerant initialisation and measurement. The intuition behind this behaviour is that $Y$ basis states are susceptible to both $X$ and $Z$ logical error mechanisms.

The most realistic scenario is that there are multiple terms in the LER expression corresponding to different classes of errors \cite{fowler2012surface}, for example single-qubit and two-qubit gates have different values for their corresponding physical error rates, hence different rates of accumulation. In this case, the ansatz could be written as:
\begin{equation} \label{eq:ler_ansatz}
    y(d) = A + B_{1q} e^{-C_{1q} d} + B_{2q} e^{-C_{2q} d}
\end{equation}

In general, we expect that adding more terms in the ansatz (which make the ansatz more expressive) will include more possible error mechanisms and will lead to a better fit. This is the functional form that will be used to obtain numerical results in Section \ref{sec:results}.

\subsection{Adapting Richardson extrapolation}

Tailoring the general Richardson extrapolation setting from Eq. \eqref{eq:richardson_yk} to our case, $\{\lambda_k\}$ are simply the inverses of the distances $\{1 / d_k\}$, since higher distances are analogous to lower levels of induced noise. Therefore, they satisfy the relation that the zero noise case corresponds to the limit of infinite distance as $\lim_{d \to \infty} 1 / d = 0$. As such, we obtain:
\begin{equation}
    E_\text{Rich}(0) = c_0 = \sum_{k = 1}^m y_k \prod_{i \neq k} \frac{d_k}{d_k - d_i}
\end{equation}

This expression will be used in Sec. \ref{sec:comparison_richardson} to compare the performance of the LER ansatz with the Richardson ansatz.

\subsection{Practical advantage}

\begin{figure}
    \centering
    \includegraphics[width=0.65\linewidth]{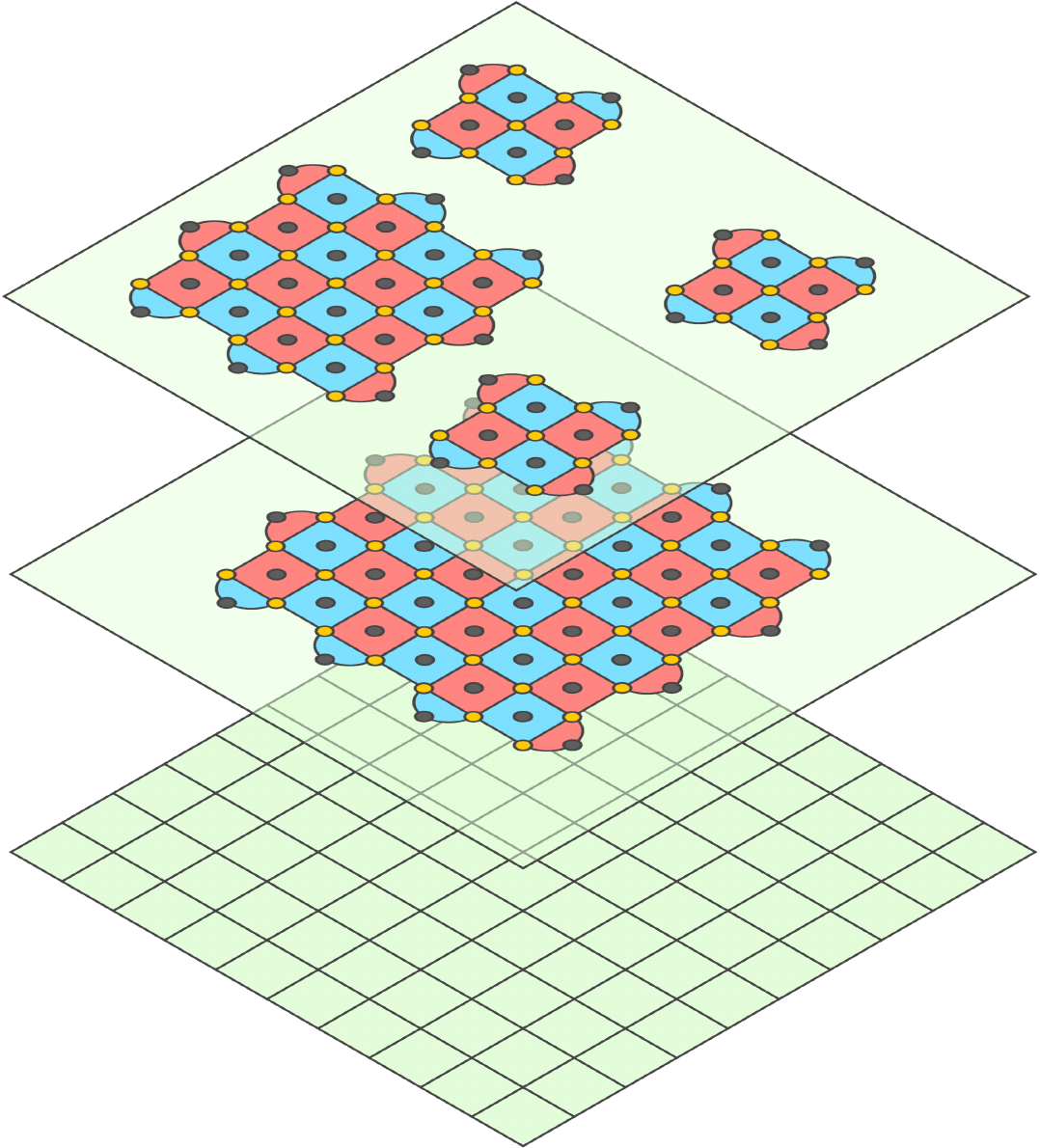}
    \caption{Illustration of how rotated surface code patches of different distances can be arranged in parallel on a device with fixed qubit resources. For example, the physical qubits can be used for either one copy of the $d=7$ code, or one copy of the $d=5$ code and three copies of the $d=3$ code.}
    \label{fig:overlapping_surface_codes}
\end{figure}

From a practical point of view, our method benefits from the advantage of being parallelisable on hardware. Specifically, if the limited number of physical qubits supports the implementation of a code with maximum distance $d_\text{max}$, we can execute a few instances of the code with distances lower than $d_\text{max}$ at the same time, as shown in Figure \ref{fig:overlapping_surface_codes}. While the Figure only shows the patches used in a memory experiment, a similar arrangement could be obtained when logical operations are involved, since the spatial size of the computation scales with the code distance.

\section{Bypassing non-stabiliser simulation}
\label{sec:bypassing_methods}

Stabiliser states, generated by circuits comprising only of Clifford gates, can be simulated classically in polynomial time. However, non-trivial states that can be prepared by a quantum computer require an extended, universal gate set that also includes non-Clifford gates, such as the $T$ gate, which makes them exponentially hard to simulate classically. In this work, we employ two methods to emulate the effect of having a non-Clifford gate while still using an efficient stabiliser simulator. This provides a more challenging test of the capabilities of IDE, as it targets non-trivial expectation values (that are not 0 or $\pm 1$).

\subsection{Method 1: Quantum process tomography}

The first idea is to use quantum process tomography (QPT) to estimate the effective logical noise channel that is implemented through a noisy, circuit-level simulation. Once we obtain the numerical description of the channel in the form of its process matrix, we can use it in conjunction with any input state (vector) to calculate arbitrary noisy expectation values. Ultimately, the process matrix is obtained by combining the expectation values of measurements and state initialisations in various Pauli bases.

To obtain the process matrix in the single qubit case, following the notation in \cite{chuang1997prescription}, we need to calculate:
\begin{align}
    \rho'_1 &= \mathcal{E}(\ket{0}\bra{0}) \\
    \rho'_4 &= \mathcal{E}(\ket{1}\bra{1}) \\
    \rho'_2 &= \mathcal{E}(\ket{+}\bra{+}) - i\mathcal{E}(\ket{+i}\bra{+i}) - (1 - i)(\rho'_1 + \rho'_4) / 2 \\
    \rho'_3 &= \mathcal{E}(\ket{+}\bra{+}) + i\mathcal{E}(\ket{+i}\bra{+i}) - (1 + i)(\rho'_1 + \rho'_4) / 2
\end{align}
where the $\mathcal{E}(\cdot)$ symbol denotes that we obtain the density matrix from quantum state tomography on the state that has been subjected to the noise channel $\mathcal{E}$ that we want to characterise. We then need to define the $\Lambda$ matrix such that:
\begin{equation}
    \Lambda = \frac{1}{2} \begin{bmatrix}
        I & X \\
        X & -I
    \end{bmatrix}
\end{equation}
and so we can finally express our desired process matrix $\chi$ as:
\begin{equation}
    \chi = \Lambda \begin{bmatrix}
        \rho'_1 & \rho'_2 \\
        \rho'_3 & \rho'_4
    \end{bmatrix} \Lambda
\end{equation}
With these prescriptions, we can now perform logical quantum process tomography on a single qubit.

\subsubsection{Quantum state tomography}

For the single qubit case, a quantum state can be described as \cite{nielsen2011quantum}:
\begin{equation}
    \rho = \frac{\Tr[\rho] I + \Tr[\rho X] X + \Tr[\rho Y] Y + \Tr[\rho Z] Z}{2}
\end{equation}
This means that it is enough to estimate the $X$, $Y$ and $Z$ observables to obtain a description of the state as a density matrix. From the previous section, we know how to do this for all three Pauli observables.

\begin{figure*}[t]
    \centering
    \begin{subfigure}[c]{0.32\textwidth}
        \centering
        \includegraphics[width=1.03\linewidth]{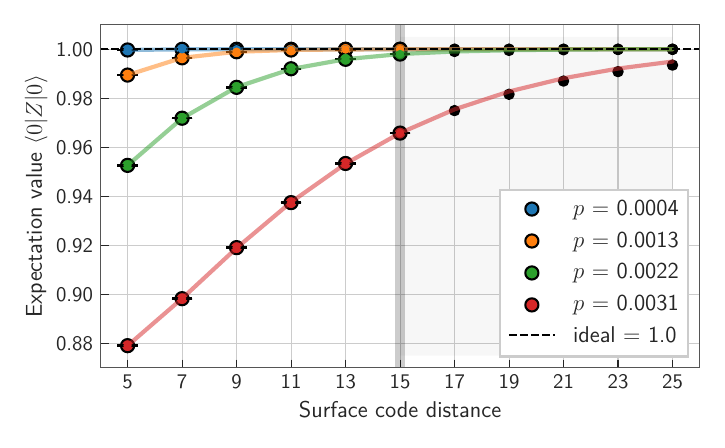}
        \caption{$\braket{0|Z|0}$ for \textit{odd} distances.}
        \label{fig:exp_val_0_Z}
    \end{subfigure}
    \hfill
    \begin{subfigure}[c]{0.32\textwidth}
        \centering
        \includegraphics[width=1.03\linewidth]{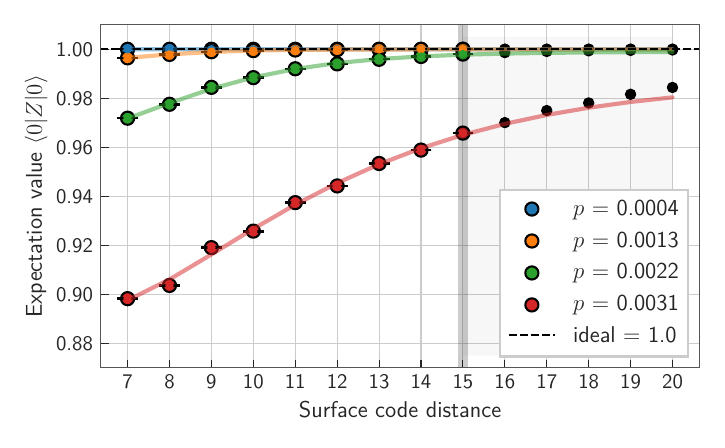}
        \caption{$\braket{0|Z|0}$ for \textit{odd and even} distances.}
        \label{fig:exp_val_0_Z_odd_even}
    \end{subfigure}
    \hfill
    \begin{subfigure}[c]{0.32\textwidth}
        \centering
        \includegraphics[width=1.03\linewidth]{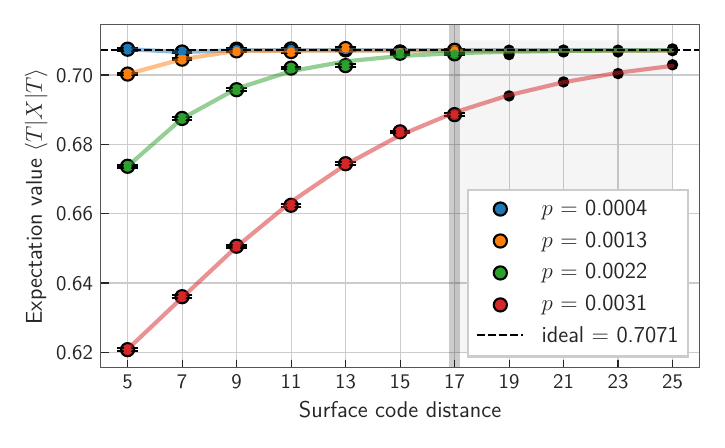}
        \caption{$\braket{T|X|T}$ for various distances.}
        \label{fig:exp_val_T_X}
    \end{subfigure}
    \caption{Expectation values as a function of surface code distance for different physical error rates $p$. The grey vertical line separates fitting data (coloured points, left) from reference data (black points, right), where the reference values are the true expectation values at those distances. The solid lines are obtained by fitting the left points to the extrapolating function ansatz. \textbf{Values closer to the horizontal black line, which represents the noiseless expectation value, are better.}}
    \label{fig:exp_vals}
\end{figure*}

\subsection{Method 2: Stabiliser decomposition}
\label{sec:non-Clifford}

A non-stabiliser state $\rho$ can be written as a sum of pure $n$-qubit stabiliser states $\{\sigma_i\}$ such that \cite{howard2017application}:
\begin{equation}
    \rho = \sum_{i = 1}^k x_i \sigma_i \quad \quad \textrm{with} \sum_{i = 1}^k x_i = 1
\end{equation}
which is called a stabiliser pseudomixture and is a quasiprobability decomposition since each $x_i$ can also be negative. The problem of finding the best such decomposition for an arbitrary state can be turned into a constrained optimisation problem for the robustness of magic, $\mathcal{R}(\rho)$, such that:
\begin{equation}
    \mathcal{R}(\rho) = \min ||x||_1 \hspace{0.5cm} \textrm{subject to} \hspace{0.1cm} Ax = b
\end{equation}
where $\min ||x||_1 = \sum_i |x_i|$, $b_i = \Tr[\rho P_i]$, $A_{j, i} = \Tr[P_j \sigma_i]$ and $P_j$ is the $j$-th Pauli operator. As an application, let us consider the decomposition of the state $\ket{T} = (\ket{0} + e^{i \pi / 4} \ket{1}) / \sqrt{2}$ which, following \cite{howard2017application}, is:
\begin{equation}
    A = \begin{pmatrix}
        1 & 1 & 1 & 1 & 1 & 1 \\
        1 & -1 & 0 & 0 & 0 & 0 \\
        0 & 0 & 1 & -1 & 0 & 0 \\
        0 & 0 & 0 & 0 & 1 & -1 \\
    \end{pmatrix} \quad \textrm{and} \quad b = \begin{pmatrix}
        1 \\
        \frac{1}{\sqrt{2}} \\
        \frac{1}{\sqrt{2}} \\
        0
    \end{pmatrix}
\end{equation}
which gives the solution $x = (\sqrt{2}, 0, 1, 1 - \sqrt{2}, 0, 0) / 2$.

Thus, by solving the constrained optimisation problem, we can find decompositions of the non-stabiliser input state, such that non-trivial expectation values will be obtained as linear combinations of efficient Clifford simulations with stabiliser input states.

The two methods presented allow us to bypass the need for direct non-Clifford simulation.

\section{Numerical Simulations}
\label{sec:results}

We perform memory experiments using the rotated surface code, and we execute $3d$ rounds of stabiliser measurements to emulate a longer computation, rather than just the canonical $d$ rounds required to correct measurement errors. The purpose of the numerical simulations is to validate the proposed ansatz, as a full analytical analysis for circuit-level noise is impossible due to the non-linearity of the LM solver.

\paragraph{Noise model}

We choose the circuit-level noise model \texttt{SI1000} introduced in \cite{gidney2021fault}. For the rotated surface code with a MWPM decoder, we observe a threshold of around 0.45\% in our simulations.

\paragraph{Numerical setup}

We use the software package \texttt{Stim} \cite{gidney2021stim} as an efficient stabiliser simulator. For decoding, we use the \texttt{PyMatching} \cite{higgott2022pymatching, higgott2025sparse} package which implements a fast variant of the minimum weight perfect matching (MWPM) algorithm adapted to the quantum decoding problem. To leverage the parallel processing power of our computers for Monte Carlo sampling and decoding, we use the \texttt{Sinter} subpackage of \texttt{Stim}.

\paragraph{Fitting method}

The functional form of the ansatz chosen for the simulations is the double-exponential form in Eq. \eqref{eq:ler_ansatz}. For the fitting procedure of our ansatz to the data, we use the non-linear Levenberg-Marquardt (LM) least squares method \cite{more2006levenberg}, as implemented in \texttt{SciPy}.

\subsection{Memory experiments}
To test our hypothesis, we consider the setting of QEC memory experiments. In this case, no logical gates are applied, and the overall circuit has the effect of logical identity. $3d$ rounds of stabiliser measurements are performed to correct for measurement errors. In all the following cases, state preparation and final measurements are implemented noiselessly, so the noisy operations only affect the QEC cycles. Most of our investigations begin at $d = 5$ to avoid possible finite-size effects present at $d = 3$ and 4, given that the fitting procedure would be particularly sensitive to them. In our simulations we use up to 100,000,000 shots for distances up to $d = 15$, usually used for fitting, and up to 2,000,000 shots for distances $d = 16 - 26$, usually used as reference.

\subsubsection{Stabiliser input state: the $\ket{0}$ state}

\paragraph{Odd distances}

The results of the extrapolation in the case of the $Z$ observable on the $\ket{0}$ state are presented in Figure~\ref{fig:exp_val_0_Z}. Unsurprisingly, the expectation value gets further away from the noiseless value (1.0) with increasing the physical error rate $p$. The points up to distance 15 are used for fitting the ansatz, and the points from distance 17 onwards are predicted using the fitted formula. In this case, the extrapolation works well because the fitting algorithm used in this particular case is a bounded version of non-linear least squares, so the expectation value is forced to be in the $[-1, +1]$ interval. Therefore, to show a fair accuracy test of the method, it is essential to consider a setting where the target expectation value is not $\pm 1$. The $\pm 1$ values also have a special status because they are least affected by shot noise, as technically a single shot is enough to perfectly calculate them under noiseless circumstances, while for other expectation values more samples are needed to achieve a desired accuracy, with an expectation value of 0 being the most expensive to determine.

\paragraph{Comparison between odd and even distances}

\begin{figure*}[t]
    \centering
    \begin{subfigure}[c]{0.496\textwidth}
        \centering
        \includegraphics[width=1.02\linewidth]{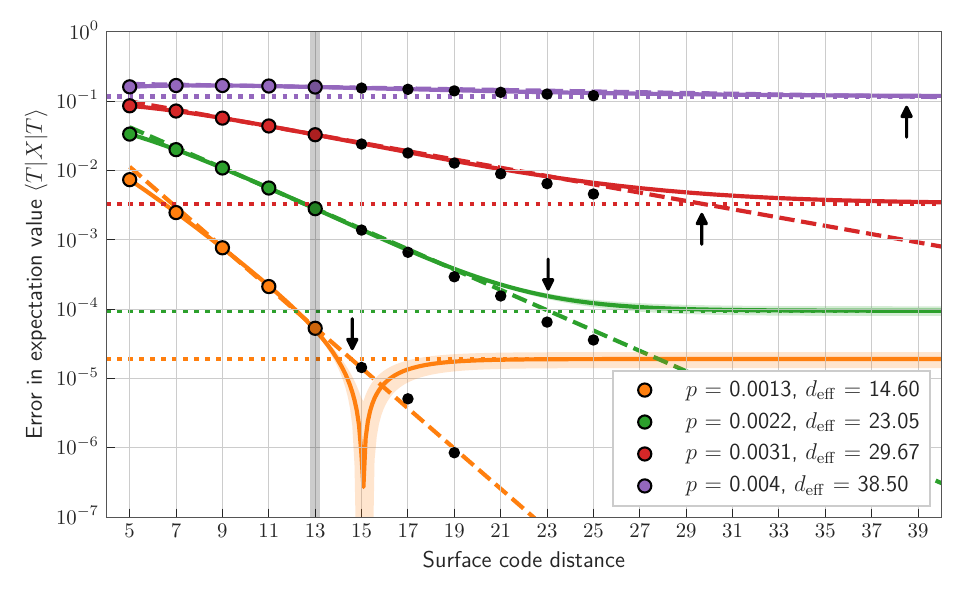}
        \caption{}
        \label{fig:T_log}
    \end{subfigure}
    \hfill
    \begin{subfigure}[c]{0.496\textwidth}
        \centering
        \includegraphics[width=1.02\linewidth]{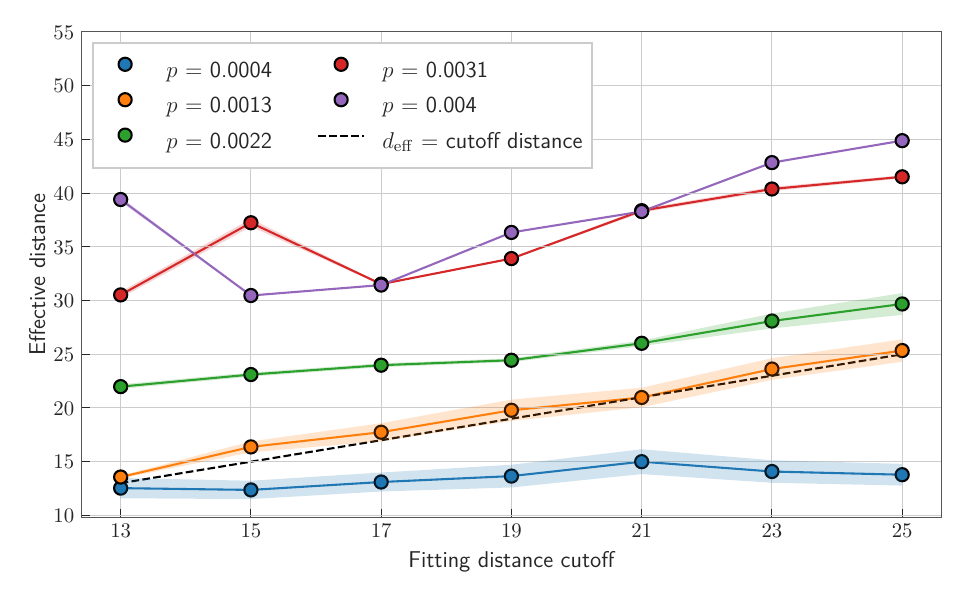}
        \caption{}
        \label{fig:T_cutoff}
    \end{subfigure}
    \caption{IDE performance for $\braket{T|X|T}$. \textbf{(a):} Absolute errors in expectation values for several values of the physical error rate $p$ and different distances $d$ of the surface code. The grey vertical line separates fitting data (coloured points, left) from reference data (black points, right). The solid lines are obtained by fitting the left points to the extrapolating function ansatz, and the shaded regions around them represent the error bars. The dotted, horizontal lines are the errors in the extrapolated expectation values at infinite distance. The dashed lines represent the linear fit to the exact data points left of the grey line. The $d_\text{eff}$ values, indicated by black arrows and listed in the legend, are calculated over multiple bootstrap resampling trials as the effective distance at which the IDE error matches the linear fit error. \textbf{Lower $(\downarrow)$ is better.} \textbf{(b):} Effective distance of the IDE value as a function of the cutoff fitting distance for different values of the physical error rate $p$. Points above the dashed $y = x$ line indicate that IDE yields an improvement over a single experiment at the cutoff distance. \textbf{Higher $(\uparrow)$ is better.}}
\end{figure*}

It is also interesting to consider the case of both odd and even distances, shown in Figure~\ref{fig:exp_val_0_Z_odd_even} for the $Z$ observable on the $\ket{0}$ state. Here, we can observe that fitting a single line through all points does not give a very good fit, while fitting the odd and the even points separately would give two separate good fits, as shown in Table \ref{tab:r2score}. Therefore, we can have more confidence in our extrapolated values by being able to compare the two cases and checking if they are in agreement. This behaviour stems from the fact that, even though distances $2k + 1$ and $2k + 2$ correct the same number of errors, the even distance code has more qubits, which increases the redundancy and hence the robustness of the system, leading to lower logical error rates in the even case. More specifically, there is an extra $Z$ stabiliser being added by going from an odd distance to the next even distance, while the number of $X$ stabilisers stays the same, leading to a mismatch in the two distances: $d_Z \neq d_X$. A related phenomenon has been observed when comparing unrotated and rotated versions of the surface code for the same distance \cite{o2025compare}, where the increased number of physical qubits in the unrotated version improves the robustness and the logical error rate of the code.

\begin{figure*}[t]
    \centering
    \begin{subfigure}[c]{0.50\textwidth}
        \centering
        \includegraphics[width=1.02\linewidth]{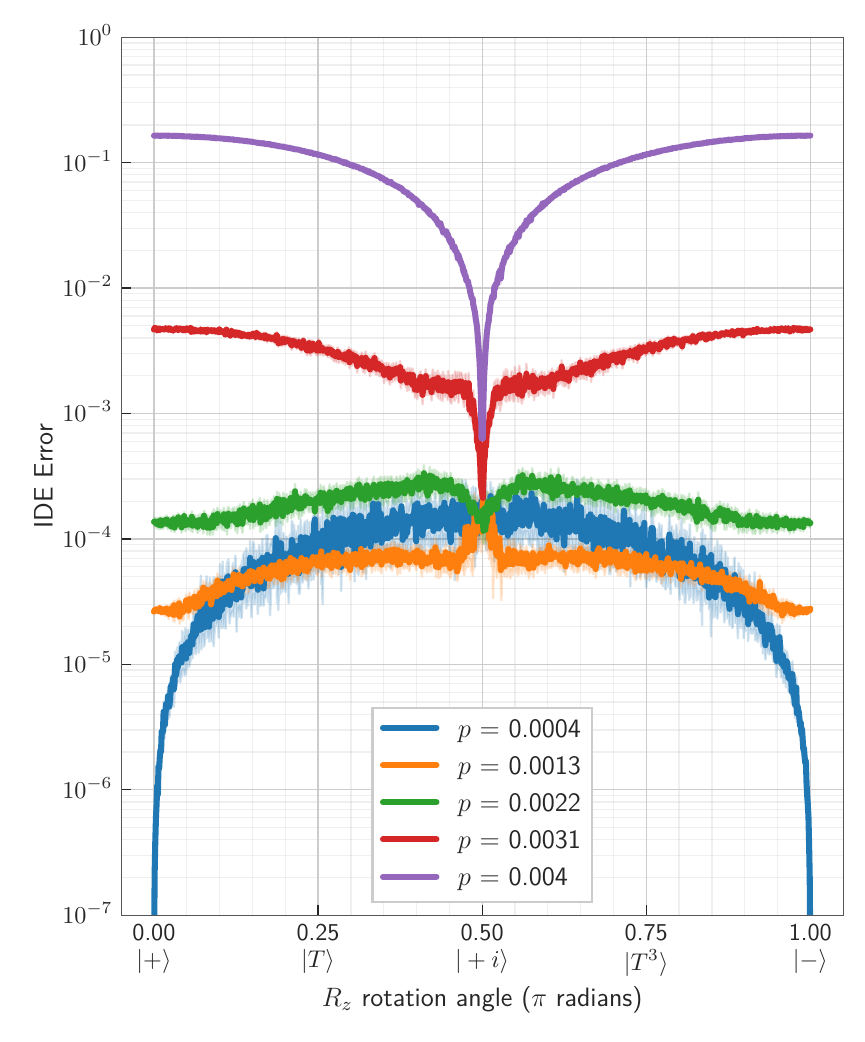}
        \caption{Error in the estimated IDE value when compared to the noiseless value. \textbf{Lower $(\downarrow)$ is better.}}
        \label{fig:ide_angles_error}
    \end{subfigure}
    \hfill
    \begin{subfigure}[c]{0.48\textwidth}
        \centering
        \includegraphics[width=1.02\linewidth]{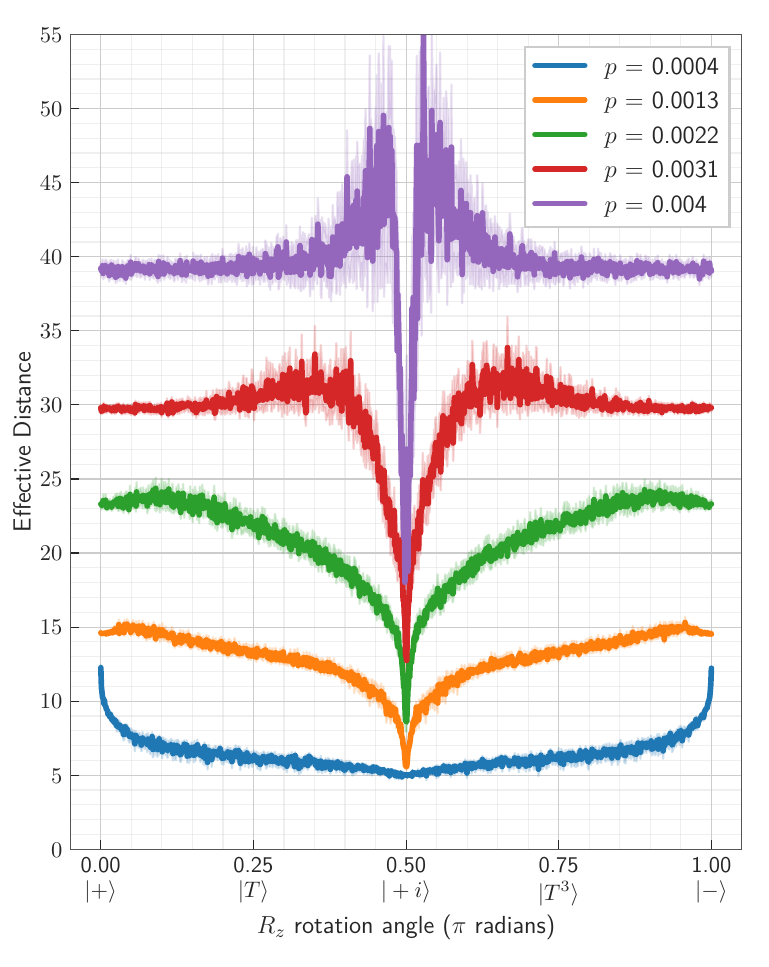}
        \caption{Effective distance of the estimated IDE value. \textbf{Higher $(\uparrow)$ is better.}}
        \label{fig:ide_angles_deff}
    \end{subfigure}
    \caption{IDE performance for the $X$ observable evaluated on non-stabiliser states in the $XY$-plane for several physical error rates, as a function of the $R_z$ rotation angle preparing the non-stabiliser state. For both subplots, the exact point at $\theta = 0$ is omitted because of numerical instability, while its label is retained on the axis for reference. The cutoff distance for fitting is $d = 13$.}
    \label{fig:ide_angles}
\end{figure*}

\subsubsection{Non-stabiliser input state: the $\ket{T}$ state}

Let us now consider one of these non-special cases that do not have an expectation value of $\pm 1$. Here, we are considering an $X$ observable on the $\ket{T}$ state, which should have an expectation value of about 0.7071. This is a relevant setting for general quantum computations, as the $\ket{T}$ state is commonly used as a resource for the non-Clifford $T$ gate, which allows universal computation. The results of the extrapolation are presented in Figure~\ref{fig:exp_val_T_X}. The points up to distance 17 are used for fitting the ansatz, the points beyond distance 17 are used as reference, and the solid lines are obtained using the fitted formula. We can clearly see that the extrapolation is not performing as well as in the case of the $\ket{0}$ state, as shown quantitatively in Table \ref{tab:r2score}. In this situation, the expectation values are calculated using the sum of stabiliser states method from Section \ref{sec:non-Clifford}. Since each expectation value is now a sum of scaled expectation values from a few simulations, the variance increases and we have a less smooth trend to observe.

We now introduce a quantity called the effective distance $d_\text{eff}$. Assume that we fit a linear function of the distance to the errors of the data points used for the fitting of the ansatz. This will be our reference function, which is advantageous in terms of simplicity and more reliable than our ansatz as it does not contain the exact expectation value as an additional parameter to be fitted. The distance at which this linear function intersects the extrapolated IDE value is defined as the effective distance $d_\text{eff}$. The effective distance is therefore a measure of how well the extrapolation performs, showing that by fitting data points up to a maximum distance $d$, we obtain similar performance to running a single experiment at distance $d_\text{eff}$. The higher the effective distance, particularly compared to the cutoff distance used for the ansatz fitting, the better IDE works. A related analysis using a linear function of the distance has been used in the literature to extrapolate to a low logical error rate limit and estimate the ``teraquop footprint'' \cite{gidney2021fault}.

Figure \ref{fig:T_log} shows more clearly the errors on the extrapolated values on the logarithmic scale. We also show the uncertainty in the extrapolated ansatz (shaded regions around solid lines) through bootstrap resampling with 1,000 trials. The errors in the estimate increase and the effective distance decreases as the physical error rate decreases. With a cutoff distance of $d = 13$ for the ansatz fitting, the method is shown to work at all the 4 considered physical error rates. The method performs worse at lower physical error rates due to shot noise relative to final logical error, which makes its use more practical when the signal is neither dominated by shot noise, for the case of low logical error rate, nor by circuit noise, for the case of very high logical error rate. We note the special case of the highest physical error rate at $p = 0.004$, which is very close to threshold for this experiment, and whose LER improvement with increasing distance is rather small. We also note the feature present in the extrapolated solid line for $p = 0.0013$. For the higher physical error rates, the extrapolated curves always plateau below the ideal expectation value (as seen in Figure \ref{fig:exp_vals}), while for this particular physical error rate, the extrapolated curve passes the exact expectation value and plateaus just above it. This is why the plot of the absolute value of this error shows the error first decreasing sharply, and then increasing to a plateau.

Figure \ref{fig:T_cutoff} shows the effective distance as a function of the cutoff distance used for ansatz fitting. The $d_\text{eff}$ values in Figure \ref{fig:T_cutoff} are calculated over 1,000 bootstrap resampling trials as the effective distance at which the IDE error rate matches the linear fit error rate. As expected, the quality of data (signal) generally increases with distance and hence the extrapolation performance, but more shots will be needed to fit higher distances. Although the increasing trend is apparent, it is encouraging to see that IDE already shows very good improvements at cutoff distances as low as $d = 13$.

\subsubsection{Non-stabiliser input states: the $XY$-plane}

We now consider a general family of input states in the $XY$-plane of the Bloch sphere, which could be obtained by applying an $R_z(\theta)$ rotation to the $\ket{+}$ state. We use the methods presented in Section \ref{sec:bypassing_methods} to simulate memory experiments for such input states. For the $X$ observable, the noiseless expectation value varies smoothly across this family: it is $+1$ at $\theta=0$ (the $\ket{+}$ state), $0$ at $\theta=\pi/2$ (the $\ket{+i}$ state), and $-1$ at $\theta=\pi$ (the $\ket{-}$ state). Figure \ref{fig:ide_angles} therefore probes IDE across a continuous range of non-stabiliser inputs, from states with extremal expectation values at the edges to states with vanishing expectation value near the middle.

Each expectation value is estimated from samples taking values in $\{+1,-1\}$. The variance of a binomial distribution is $\text{Var}[X] = 4q(1 - q)$, where $q$ is the probability of one of the outcomes. The variance is therefore largest when the expectation value is close to $0$ ($q = 0.5$), and smallest when the expectation value is close to $\pm 1$. As a result, the central region of Figure \ref{fig:ide_angles} is intrinsically harder to estimate than the edges. This provides a useful stress test of IDE, since the quality of the extrapolation is affected not only by the underlying logical error rate, but also by the state-dependent sampling variance.

In Figure \ref{fig:ide_angles_error} we show the absolute error of the IDE estimate relative to the noiseless expectation value, while Figure \ref{fig:ide_angles_deff} shows the corresponding effective distance. The overall message is that, while the performance is not uniform, IDE remains effective across a broad range of input states, not only for stabiliser states and the $\ket{T}$ state considered earlier. Across most of the angle landscape, the extrapolated errors and effective distances tend to be larger for higher physical error rates, with the exception of the lowest physical error rate, $p=0.0004$, where the limited number of shots leads to noticeably larger estimation error. With a cutoff distance of $d=13$, this is also the only physical error rate for which IDE does not yield an effective-distance improvement.

Figure \ref{fig:ide_angles_error} reveals several competing effects. At the edges of the angle range ($\theta$ near $0$ or $\pi$), the noiseless expectation value is $\pm 1$, and the absolute IDE error is dominated by logical errors rather than shot noise. Here, higher physical error rates lead to larger IDE errors, as expected. Near the centre ($\theta \approx \pi/2$), the noiseless expectation value is close to $0$, and the IDE error is instead dominated by shot noise in the input data, since the variance of the estimator is maximal. In this regime, lower physical error rates, which produce fewer logical errors and hence smaller deviations from the noiseless value, are more susceptible to shot noise, leading to larger IDE errors. This explains the crossover visible between $p = 0.0022$ and $p = 0.0013$: at the centre of the landscape, $p = 0.0013$ transitions from having the lowest IDE error to the highest.

In Figure \ref{fig:ide_angles_deff}, the separation between the different physical error rates is much cleaner, and the trend is more consistent for all of them, with the effective distances at the edges being higher than in the middle. However, we still see the interplay between the physical error rates and the variance of the EV, as some peaks become apparent around the middle for the higher physical error rates. Even when the absolute IDE error in Figure \ref{fig:ide_angles_error} is not minimal at the edges, Figure \ref{fig:ide_angles_deff} shows that the extrapolation can still provide a larger gain there relative to the unmitigated estimate.

Despite the multi-step procedure involved, decomposing each non-stabiliser state into stabiliser components that are individually simulated, fitting the ansatz at each angle, and computing the effective distance, the method shows stable, smoothly varying performance across the full angle landscape. This uniformity holds across the full range of physical error rates considered, including the regimes where shot noise and logical noise contribute differently to the overall IDE error.

\begin{figure}
    \centering
    \includegraphics[width=1.03\linewidth]{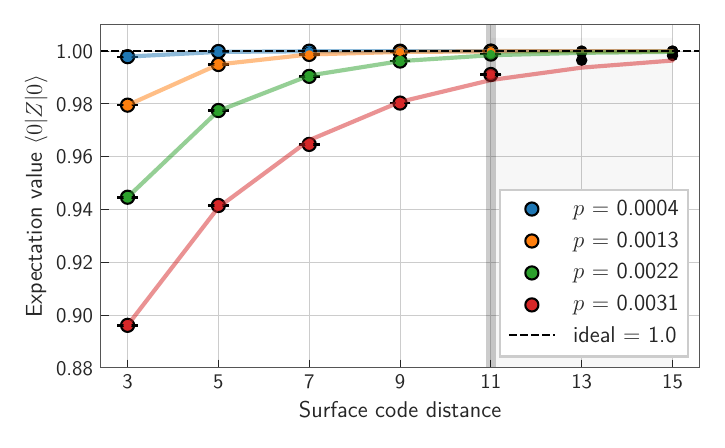}
    \caption{Expectation values $\braket{0|Z|0}$ versus surface code distance for several physical error rates $p$, for the case of spatially inhomogeneous noise. The grey vertical line separates the points used for fitting (coloured, left) from held-out reference points (black, right). \textbf{Values closer to the horizontal black line, which represents the noiseless expectation value, are better.}}
    \label{fig:exp_val_spatial_variability}
\end{figure}

\begin{figure*}[t]
    \centering
    \begin{subfigure}[c]{0.49\textwidth}
        \centering
        \includegraphics[width=1.02\linewidth]{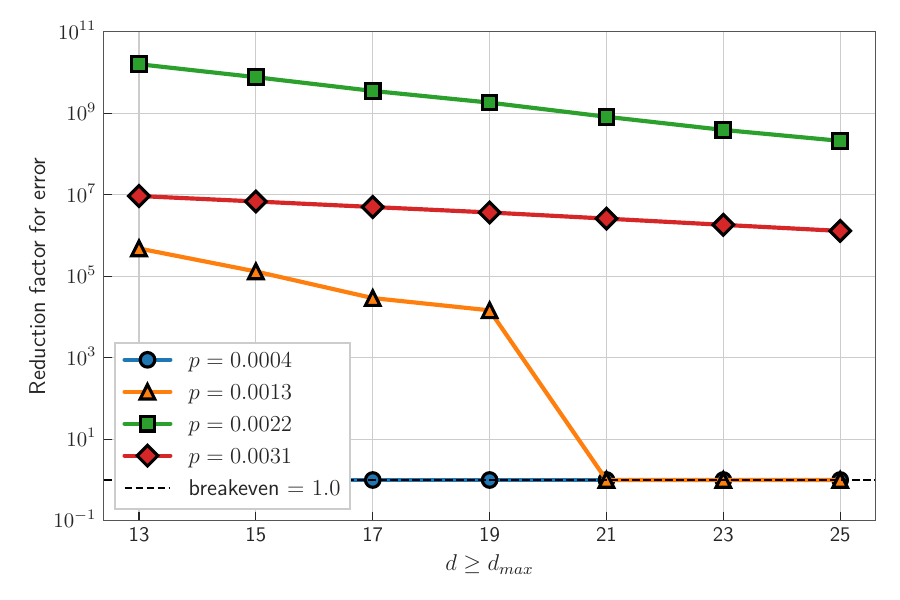}
        \caption{$\braket{0|Z|0}$. Distances up to $d_\text{max} = 13$ were used for fitting.}
        \label{fig:improvement_0}
    \end{subfigure}
    \hfill
    \begin{subfigure}[c]{0.49\textwidth}
        \centering
        \includegraphics[width=1.02\linewidth]{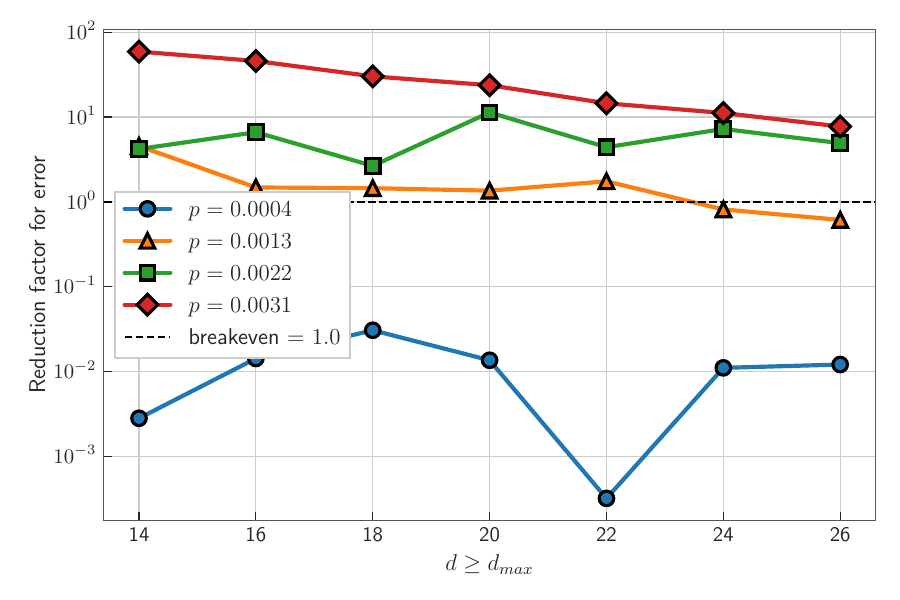}
        \caption{$\braket{T|X|T}$. Distances up to $d_\text{max} = 14$ were used for fitting.}
        \label{fig:improvement_T}
    \end{subfigure}
    \caption{Reduction factor of the error in the expectation value for different values of the physical error rate $p$. Each point on the horizontal axis corresponds to a single QEC experiment at that distance, and the data points show the factor by which IDE reduces the error relative to that experiment. \textbf{Higher $(\uparrow)$ is better.}}
\end{figure*}

\subsubsection{Robustness to spatial variations in noise}

As a highly practical consideration, we highlight that real quantum devices suffer from fabrication defects that lead to the performance of different qubits and operations to vary. We can thus examine how robust IDE is to such inhomogeneities. In the case of spatial variations, we sample different values of noise from a normal distribution centred around $p$ for each qubit or pair of qubits, depending on the operation, and maintain these across all QEC cycles. We present results of the spatial variation in Figure \ref{fig:exp_val_spatial_variability}, where we chose an inhomogeneity of 0.3, meaning that we scale the physical error rate $p$ by a normal distribution of mean 1 and standard deviation 0.3. For reference, for a $d = 13$ patch, it means that the underlying physical error rates of the qubits lie roughly in the $[0.2p, 1.7p]$ interval.

The results show that the fit is robust to inhomogeneity among physical qubits, making the method more applicable in real-world scenarios. We also observe that the resulting logical error rate is not comparable to that of the average physical error rate, but rather to that of the highest physical error rate sampled (worst-performing operation). From a practical point of view, this means that the quality of the results is always limited by the worst element of the computation.

\begin{table}[]
    \centering
    \begin{tabular}{|c|c|c|c|c|}
        \hline
        \multicolumn{1}{|c}{$R^2$} & \multicolumn{4}{|c|}{$p$} \\
        \hline
        Figure & 0.0004 & 0.0013 & 0.0022 & 0.0031 \\
        \hline
        \ref{fig:exp_val_0_Z} & 0.99999 & 0.99999 & 0.99998 & 0.99958 \\
        \ref{fig:exp_val_0_Z_odd_even} - odd & 1.0 & 0.99998 & 0.99993 & 0.99947 \\
        \ref{fig:exp_val_0_Z_odd_even} - even & 0.99992 & 0.99986 & 0.99994 & 0.99996 \\
        \ref{fig:exp_val_0_Z_odd_even} - all & 0.99579 & 0.99870 & 0.99722 & 0.99529 \\
        \ref{fig:exp_val_T_X} & 0.17995 & 0.97997 & 0.99653 & 0.99965 \\
        \ref{fig:exp_val_spatial_variability} & 0.99385 & 0.99999 & 0.99984 & 0.99770 \\
        \hline
    \end{tabular}
    \caption{$R^2$ score showing the agreement between the fitted function and the data points considered for some of the previous figures. \textbf{Higher $(\uparrow)$ is better.}}
    \label{tab:r2score}
\end{table}

\subsubsection{Goodness of fit}

We use the $R^2$ score to measure the goodness of fit for the figures displaying expectation values. The numerical values can be found in Table \ref{tab:r2score}. We observe that, sometimes, the fit degrades for lower physical error rates due to few logical errors being sampled, which increases the variance on the estimate of the EV, leading to more uncertainty in the fitting procedure. We also notice how the fit for the combined odd and even distances in Figure \ref{fig:exp_val_0_Z_odd_even} is significantly worse compared to the separate fits of the odd and even distances.

\begin{table}[]
    \centering
    \begin{tabular}{|c|c|}
        \hline
        $d \geq d_\text{max}$ & Total physical qubits \\
        \hline
        13 & 337 \\
        15 & 449 \\
        17 & 577 \\
        19 & 721 \\
        21 & 881 \\
        23 & 1057 \\
        25 & 1249 \\
        \hline
    \end{tabular}
    \caption{Number of physical qubits (data and ancilla) required for different distances of the rotated surface code.}
    \label{tab:physical_qubits}
\end{table}

\subsubsection{Reduction in error by using IDE}

In this section, we aim to quantify the improvement in the accuracy of the estimated expectation values by the improvement ratio defined as:
\begin{equation}
    IR(d, d_\text{max}) = \frac{| E_d - E^*|}{|E^\text{ext}_{d_\text{max}} - E^*|}
\end{equation}
where $E_d$ is the error-corrected expectation value calculated at that particular distance, with no QEM, $E^\text{ext}_{d_\text{max}}$ is the extrapolated (mitigated) expectation value using the error-corrected expectation values from same-parity distances up to $d_\text{max}$, and $E^*$ is the noiseless expectation value. The improvement ratio is the inverse of an important quantity that has been used in the error mitigation literature, namely the relative error of mitigation \cite{cirstoiu2023volumetric, saki2023hypothesis}.

The results are presented in Figure \ref{fig:improvement_0} for the $Z$ observable on the logical $\ket{0}$ state and in Figure \ref{fig:improvement_T} for the $X$ observable on the logical $\ket{T}$ state. When the ratio is above 1, it means extrapolation is improving the EV, which is the case for almost all the situations considered. For example, in Figure \ref{fig:improvement_T}, we can see that even with our imperfect non-stabiliser emulation, we get a $10 \times$ reduction in error by using data from experiments up to distance 14, compared to running a single QEC experiment at distance 26, which requires a chip that is roughly $4 \times$ larger (see Table \ref{tab:physical_qubits}). When the ratio drops to 0, it means that there are no sampled logical errors after QEC at those distances, which is particularly the case for high distances and low physical error rates. It is also only the case for the lowest error rate for the $\ket{T}$ state that QEM is actually detrimental.

\begin{figure}
    \centering
    \includegraphics[width=1.02\linewidth]{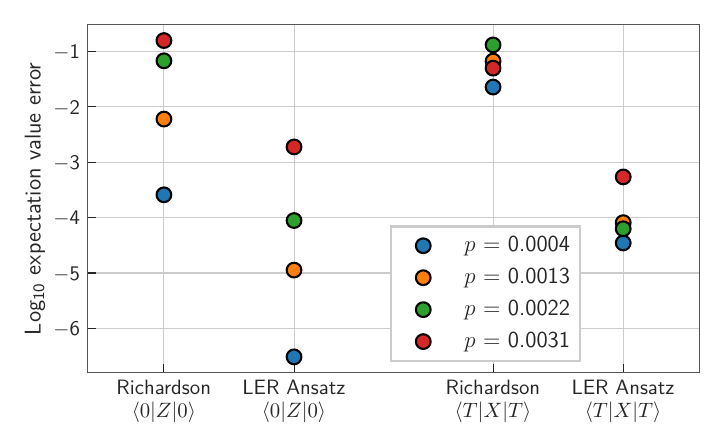}
    \caption{Logarithm of the difference (error) in expectation value between the extrapolated value and the noiseless case. These errors are compared for $\braket{0|Z|0}$ and for $\braket{T|X|T}$, for the Richardson extrapolation and the extrapolation using a logical error rate (LER) ansatz, as well as for 4 different levels of physical error rate. \textbf{Lower $(\downarrow)$ is better.}}
    \label{fig:tomography_rich_ext}
\end{figure}

\subsubsection{Comparison with the Richardson ansatz}
\label{sec:comparison_richardson}

A comparison of two different ansatze, the LER ansatz (Eq. \eqref{eq:ler_ansatz}) and the Richardson ansatz (Eq. \eqref{eq:richardson_ansatz}), is shown in Figure~\ref{fig:tomography_rich_ext}, where the underlying data points used for fitting for both methods are the original, single-distance, error-corrected expectation values. The LER ansatz outperforms the traditional Richardson extrapolation for both the $\ket{0}$ state and the $\ket{T}$ state, with logical error rates after extrapolation that are up to 3 orders of magnitude lower. For the $\ket{T}$ state, we observe an ordering of the presented points that is non-monotonic with the underlying physical error rates, caused by the non-linear accumulation of errors given by the addition of several expectation values in the simulation procedure.

\subsection{Resource savings}

Let us assume that we can run a hardware experiment at code distance $d$. The quantity $\Lambda_{d/d+2}$ represents the factor of reduction in the LER from using a code distance of $d + 2$ instead of the original $d$ \cite{google2021exponential}. If by running IDE with distances up to $d$, we obtain a factor $f$ of reduction in the LER, and we assume a constant factor of $\Lambda_{d/d+2}$, then the improvement in the effective distance of the code is:
\begin{equation}
    d_\text{eff} - d = 2 \times \frac{\log f}{\log \Lambda_{d/d+2}}
\end{equation}
Numerical examples are presented in Table \ref{tab:lambda}. The value of $\Lambda_{d/d+2} = 2$ is indicative of the NISQ regime \cite{eickbusch2025demonstration}, $\Lambda_{d/d+2} = 5$ for the EFT regime, and $\Lambda_{d/d+2} = 10$ for the late FT regime.

\begin{table}[]
    \centering
    \begin{tabular}{|c|c|c|c|}
        \hline
        $\Lambda_{d/d+2}$ & \makecell{Reduction\\LER ($f$)} & \makecell{Increase\\Distance} & \makecell{Reduction\\\# Qubits} \\
        \hline
        \multirow{3}{0.5em}{2} & $\div10$ & +6.64 & $\div3.02$ \\
        & $\div100$ & +13.29 & $\div6.13$ \\
        & $\div1000$ & +19.93 & $\div10.33$ \\
        \hline
        \multirow{3}{0.5em}{5} & $\div10$ & +2.86 & $\div1.74$ \\
        & $\div100$ & +5.72 & $\div2.68$ \\
        & $\div1000$ & +8.58 & $\div3.82$ \\
        \hline
        \multirow{3}{1em}{10} & $\div10$ & +2 & $\div1.49$ \\
        & $\div100$ & +4 & $\div2.09$ \\
        & $\div1000$ & +6 & $\div2.78$ \\
        \hline
    \end{tabular}
    \caption{Based on a range of values of $\Lambda_{d/d+2}$, and for a range of improvement factors of the LER ($f$ values), the table shows the improvements in the effective distance (additive) and in the number of physical qubits (multiplicative). The baseline distance used for the improvement in the number of physical qubits is 9. \textbf{Higher $(\uparrow)$ is better.}}
    \label{tab:lambda}
\end{table}

Let us use as a guiding example a hypothetical improvement factor of $f = 1,000$. Based on the values calculated in Table \ref{tab:lambda}, this would correspond to an improvement of +20 in the effective distance in the near-term, and an improvement of +2 in the long-term. In terms of the physical resources required, in the near-term we could get a $10\times$ reduction in the number of physical qubits required by a quantum computer to achieve a similar LER, while in the long-term we would get closer to a $3\times$ reduction. Additionally, since the number of rounds of stabiliser measurements scales with $d$, the time of the computation is also reduced by a factor of $1.5 - 3 \times$, which results in an overall space-time volume saving of roughly $4.5 - 30 \times$. Even in the more pessimistic case of only a $10\times$ improvement in the LER by using IDE, we can still expect a space-time volume saving of $2 - 5 \times$ between the near- and long-term regimes. It is worth pointing out that the use of IDE is available for any device in its current, unmodified form, while implementing the hardware improvements necessary to achieve similar LERs would require considerable additional costs. 

However, using data points at distances lower than the maximal distance on the device comes with its own trade-offs. First, as the distance decreases, the LER increases and eventually saturates at 50\%, beyond which the data points carry no useful information for fitting. For hardware with high $\Lambda$, this saturation is reached after only a few reductions in distance, limiting the number of usable data points for the ansatz. Second, even before saturation, the higher LER at lower distances pushes expectation values closer to zero, increasing the variance of the estimate and reducing the precision of each data point used for fitting. As such, it is conceivable that an ansatz with more parameters that should technically perform better could actually perform worse than a smaller ansatz due to the additional lower-distance data points being either saturated or too noisy. Investigating the interplay between these effects for realistic circuit sizes and $\Lambda$ values is an interesting direction for future work.

\section{Conclusion}
\label{sec:conclusion}

In this paper, we have investigated an adaptation of the QEM method of ZNE to the QEC setting, which comes with a more controlled scaling variable due to the smooth variation of the LER with the logical code distance. We have shown that IDE can outperform the unmitigated case across a broad regime of trivial and non-trivial expectation values, with realistic, circuit-level noisy simulations for the surface code, a leading candidate for practical QEC implementations. We have demonstrated this performance for a logical qubit, and we have also shown how the technique is robust to spatial variations in the noise.

We find resource savings for physical error rates in the range $p=0.22\%$ to $0.4\%$.
At $p = 0.22\%$, we estimate a reduction in qubit overhead up to $\approx 3\times$, whereas for $p = 0.4\%$ this rises up to $\approx 9\times$.
We note that these savings hold provided the absolute value of the expectation value is sufficiently large (in this case around 0.3 or larger), and we see non-zero improvements for a broader regime of expectation values larger than 0.05. These benefits come at the cost of an additional sampling overhead typical of error mitigation methods.

We remind the reader that, even though infinite distance extrapolation has the chance of overshooting the expectation value, we can still extrapolate to a finite distance with higher probability of obtaining an improvement. All these results make the method appealing for EFT implementations rather than the late-FT regime, and further demonstrate that QEM can be a useful addition to QEC calculations below the physical error threshold. We would also like to point out, as it is also mentioned in \cite{wahl2023zero}, that running the experiments for different distances can be parallelised on the same chip, where it should be the case that separate patches can be used for smaller distances within the same quantum computer.

In contrast to earlier work \cite{wahl2023zero}, we test the hypothesis that the way that logical noise channels scale as a function of distance is compatible with error extrapolation, and we employ circuits beyond the simplest and atypical case of a $+1$ expectation value and consider observables of non-stabiliser states.  By performing a full circuit-level noise simulation of end-to-end QEC memory experiments for the surface code, we are able to provide a more in-depth justification of the functional form of the ansatz. The circuit-level noise model for which the logical error rate formula is derived \cite{fowler2012surface} is a simpler choice compared to more recent variants, such as \texttt{SI1000}, as it can only account for fewer error mechanisms.

We would like to emphasise that the method that we investigated is not particular to the surface code, but this was just chosen as an illustrative example due to its practicality and smooth variation in distance. These small jumps in distance for consecutive members of the code family can be seen for topological codes. Possible directions for future work include finding more rigorous analytical guarantees for the performance of the technique, together with investigating its applicability to other code families. In particular, it would be interesting to examine how the method would work for LDPC code families with non-smooth distance constructions, as these are classes of codes with great potential for practical implementation, for example in neutral atom architectures \cite{bluvstein2025fault}. Higher jumps in distance would also be observed in the case of concatenated codes, so it would be worth investigating if scalings of the form $d^2$, $d^3$ and so on would be unfavourable for IDE.

\acknowledgements

G.U. would like to thank Lingling Lao for useful discussions at the beginning of this project, Asmae Benhemou for detailed discussions about non-linear least squares extrapolation, Ben Criger for useful suggestions regarding ansatz selection, and Annie Ray for useful discussions and for providing feedback for this manuscript. G.U. acknowledges support from the Engineering and Physical Sciences Research Council [grant number EP/S021582/1]. O.H.
acknowledges support from the Engineering and Physical Sciences Research Council [grant number EP/L015242/1] and a Google PhD fellowship. G.U. and D.E.B. were supported by Innovate UK [grant number 10179725: QNABLE]. The authors acknowledge the use of the UCL Myriad High Performance Computing Facility (Myriad@UCL), and associated support services, in the completion of this work. The authors acknowledge the use of the open-source libraries \texttt{NumPy} \cite{2020NumPy-Array} and \texttt{SciPy} \cite{2020SciPy-NMeth}.

\bibliographystyle{alpha}
\bibliography{ref}

\end{document}